\documentclass[prd,aps,preprintnumbers, showpacs, nofootinbib,superscriptaddress,
notitlepage
]{revtex4-1}
  \usepackage{latexsym,bm,amsmath,amssymb,amsfonts,color,graphicx,mathtools,amsthm,indentfirst,float}
\definecolor{mycolor}{RGB}{13,5,255}
\usepackage[colorlinks=True, citecolor=mycolor, urlcolor=mycolor, linkcolor=mycolor]{hyperref}
\newcommand{\beq}{\begin{eqnarray}}
\newcommand{\eeq}{\end{eqnarray}}

\newcommand{\nn}{\nonumber \\}
\newcommand{\ma}{\mathrm}

\usepackage{graphicx}
\usepackage{amsmath,amsthm,amssymb}
\usepackage{color}

\begin{document}

\title{Computing the gluon Sivers function at small-$x$ }

\author{Xiaojun Yao}
\affiliation{Department of Physics, Duke University, Durham, North Carolina 27708, USA}

\author{Yoshikazu Hagiwara}
\affiliation{ Department of Physics, Kyoto University, Sakyo-ku, Kyoto 606-8502, Japan}

\author{Yoshitaka Hatta}
\affiliation{Physics Department, Brookhaven National Laboratory, Upton, New York 11973, USA}

\begin{abstract}
We compute the gluon Sivers function $f_{1T}^{\perp g}(x,k_\perp)$ of the transversely polarized nucleon at small-$x$ by exploiting the known connection between the dipole gluon Sivers function and the Odderon. We numerically solve the evolution equation for the Odderon both in the linear and nonlinear regimes. While we find that the $x$ and $k_\perp$ dependences of the Sivers function do not factorize as a result of the quantum evolution, factorization breaking is not numerically significant, and is much milder than what one expects in the case of unpolarized TMDs.  We also point out the possibility that, due to the presence of a node in the Sivers function, single spin asymmetry for open charm production in semi-inclusive deep inelastic scattering flips signs as the transverse momentum of  D-mesons is varied. This can be tested at the future Electron-Ion Collider. 
\end{abstract}
\maketitle

\section{Introduction}

The Sivers function \cite{Sivers:1989cc} is arguably the most studied transverse momentum dependent (TMD) distribution of the nucleon.  It has been originally introduced by Sivers to explain large transverse single spin asymmetries (SSA) observed in polarized $pp$ collisions $p^\uparrow p \to hX$. Nonvanishing parton transverse momentum $k_\perp$ allows for a spin-orbit correlation of the form $k_\perp \times S_\perp$ which leads to the left-right asymmetry of parton distributions with respect to the spin vector $S_\perp$.  It has now become a standard picture that SSA predomiantly arises from such nonperturbative initial or final state effects.     
 Moreover, what makes the Sivers function particularly interesting in phenomenological applications is its T-oddness.  The sign change of the Sivers function between semi-inclusive DIS and Drell-Yan processes \cite{Collins:2002kn} is an archetypal example of the process dependence of TMD factorizations, and has been a subject of intense theoretical and experimental investigations (see \cite{Adamczyk:2015gyk,Aghasyan:2017jop} and references therein). 

While most of the literature deals with the quark Sivers function $f_{1T}^{\perp q}(x,k_\perp)$, there is a growing interest in the gluon Sivers function $f_{1T}^{\perp g}(x,k_\perp)$ and its possible manifestation at RHIC and the future Electron-Ion Collider  (EIC) and AFTER@LHC  
\cite{Anselmino:2004nk,Yuan:2008vn,Zhou:2013gsa,DAlesio:2015fwo,Boer:2015pni,Boer:2016fqd,Godbole:2017fab,Goncalves:2017fkt,Zheng:2018awe,Rajesh:2018qks,Dong:2018wsp,Hadjidakis:2018ifr,Godbole:2018mmh,DAlesio:2018rnv} (see also a review \cite{Boer:2015vso}). 
On general grounds, one expects that the gluon Sivers function is an important or even dominant source of SSA for the production of high mass states such as $D$-mesons, $J/\psi$'s or dijets. It may also contribute to the SSA of light hadrons at negative $x_F$ (Feynman variable)  which is sensitive to the small-$x$ region of the transversely polarized nucleon.   However, there has been very little guidance so far, if any, from theory perspectives what should be the functional dependence of $f_{1T}^{\perp q,g}$ on $x$ and $k_\perp$. Often the $k_\perp$-dependence is assumed to be Gaussian and factorized from the $x$-dependent part $f_{1T}^{\perp }(x,k_\perp)\sim f(x)e^{-ck_\perp^2}$ in phenomenological fits.  However, apart from convenience, there does not seem to be a good, model-independent argument to justify such assumptions. For unpolarized TMDs, factorization is certainly broken in the small-$x$ region where the BFKL and saturation effects are important. In particular, in the saturation regime it will be maximally violated,  as TMDs depend on $x$ and $k_\perp$ through the ratio $k_\perp/Q_s(x)$ with $Q_s(x)\sim 1/x^\alpha$ being the saturation momentum. Yet, for polarized TMDs such as the Sivers function, the situation is unclear due to the lack of first principle approaches. 

Recently, however, an interesting connection between the gluon Sivers function and the so-called Odderon exchange has been elucidated \cite{Zhou:2013gsa,Szymanowski:2016mbq,Boer:2015pni}. This paves the way for a numerical evaluation of the gluon Sivers function. Since the Odderon evolves leading logarithmically at small-$x$ and the corresponding  Regge intercept is exactly unity  \cite{Bartels:1999yt},  one can expect that the gluon Sivers function survives at small-$x$.   In this paper we test this possibility by  solving the small-$x$ evolution equation for the Odderon both in the linear and nonlinear regimes. We then discuss the properties of the solution and their implications in phenomenology, with particular emphasis on the factorization of $x$ and $k_\perp$ dependences. 

The paper is organized as follows. In Section~\ref{sec:sivers_odderon}, we review the relation between the gluon Sivers function and the Odderon. Then in Section~\ref{sec:solution}, we solve the small-$x$ evolution equation of the Odderon, and using this solution we compute the gluon Sivers function.  Phenomenological implications of our results on the SSA in open charm production are  discussed in Section IV.  Section~\ref{sec:conclusion} is devoted to conclusions.

\section{Gluon Sivers function and Odderon}
\label{sec:sivers_odderon}
The gluon sivers function of the nucleon $f_{1T}^\perp$ is defined by the matrix element
\beq
\frac{1}{xP^+}\int \frac{dz^- d^2z_\perp}{(2\pi)^3} e^{-ixP^+z^- +ik_\perp \cdot z_\perp} \langle PS_\perp|2{\rm Tr}[ F^{+i}(z^-,z_\perp)U^{[\pm]}F^{+i}(0)U^{[\pm]}] |PS_\perp\rangle 
\nn
 = f_1^{[\pm \pm]}(x,k_\perp^2) -\frac{k_\perp \times S_\perp}{M} f_{1T}^{\perp [\pm \pm]}(x,k_\perp^2)\,,\label{left}
\eeq
 where $M$ is the nucleon mass and  $k_\perp \times S_\perp \equiv \epsilon^{ij}k^i_\perp S_\perp^j$ with $|S_\perp|=1$. $f_1(x,k_\perp^2)$ is the unpolarized gluon TMD normalized such that $\int d^2k_\perp f_1(x,k_\perp^2)=G(x)$ is the usual collinear gluon PDF.\footnote{Our normalization of $f_1$ and $f_{1T}^\perp$ differs from that of \cite{Boer:2015pni} by a factor of 2, see (2) of \cite{Boer:2015pni} and sum over the indices $\mu\nu$. }   The trace is in the fundamental representation (e.g., $F^{+i}=F^{+i}_at^a$) with the normalization ${\rm Tr}(t^at^b)=\frac{\delta^{ab}}{2}$. 
 
  $U^{[\pm]}$ is the lightlike, staple-shaped Wilson line connecting the points $[z,0]$ via light-cone infinity $x^-=\pm \infty$. The choice $[++]$ or $[--]$ corresponds to the Weisz$\ddot{{\rm a}}$cker-Williams (WW) Sivers function which has the property $f_{1T}^{\perp [++]}=-f_{1T}^{\perp [--]}$ due to $PT$ symmetry. This is the gluonic counterpart of the sign change of the quark Sivers function, and  can be tested experimentally \cite{Boer:2016fqd,Zheng:2018awe}. On the other hand, the choice $[+-]$ or $[-+]$ gives the dipole Sivers function. Again, from $PT$ symmetry it follows that  $f_{1T}^{\perp[+-]} = -f_{1T}^{\perp [-+]}$. 
 
 We shall be interested in the small-$x$ behavior of $f_{1T}^\perp$. As observed in \cite{Boer:2015pni}, the WW Sivers function is suppressed at small-$x$. On the other hand, the dipole Sivers function   develops leading logarithms $\alpha_s \ln 1/x$ and therefore it has a chance to survive in the small-$x$ region. Indeed,   neglecting $x$ in the exponential factor of (\ref{left}), we can approximate the left hand side as, for the component $[+-]$, 
 \beq
 \frac{4N_ck_\perp^2}{xg^2} \int \frac{d^2x_\perp d^2y_\perp}{(2\pi)^3} e^{ik_\perp\cdot (x_\perp-y_\perp)}S(x_\perp,y_\perp)\,, \label{s}
 \eeq
 where $S$ is the so-called dipole S-matrix  
\beq
S(x_\perp,y_\perp) = \frac{ \langle PS|\frac{1}{N_c}{\rm Tr}U^\dagger(x_\perp)U(y_\perp) |PS\rangle}{\langle PS|PS\rangle}\,, \label{mat}
\eeq
with
\beq
U(y_\perp)=P\exp\left(-ig\int^\infty_{-\infty} dz^- A^+(z^-,y_\perp)\right)\,.
\eeq
The second term of (\ref{left}) is odd in $k_\perp$, so the Sivers function is related to the imaginary part  of the S-matrix which is odd in $x_\perp-y_\perp=r_\perp$.
To see this more explicitly, we parameterize as  
 \beq
S(x_\perp,y_\perp)=P(|r_\perp|)+iS_\perp\times r_\perp Q(|r_\perp|)\,, \label{ji}
\eeq 
 where $P$, $Q$ are real. $P$ is commonly referred to as the Pomeron, whereas the imaginary part of the dipole S-matrix has been identified with the Odderon \cite{Hatta:2005as}. Since the Odderon amplitude is a scalar function and odd in $r_\perp$, it vanishes   unless there is a reference vector other than $r_\perp$. For the transversely polarized nucleon, the transverse spin $S_\perp$ provides such a vector.\footnote{Another example is that if one considers the nonforward amplitude $\langle P'|{\rm Tr}U_x^\dagger U_y|P\rangle$, the momentum transfer $\Delta_\perp = P'_\perp -P_\perp$ provides a reference vector. This is equivalent to consider the dipole S-matrix as a function also of the impact parameter $b_\perp=\frac{x_\perp+y_\perp}{2}$. The Odderon term has the structure $b_\perp \cdot r_\perp$ in this case \cite{Kovchegov:2012ga}. }   The resulting Odderon term has the structure (\ref{ji}) and has been dubbed the spin-dependent Odderon \cite{Zhou:2013gsa,Szymanowski:2016mbq}.

Now consider the Fourier transform of (\ref{ji})
\beq
\int d^2r_\perp S(x_\perp,y_\perp) e^{ik_\perp \cdot r_\perp} = \widetilde{P}(|k_\perp|) + S_\perp \times k_\perp \widetilde{Q}(|k_\perp|)\,, \label{we}
\eeq
where
\beq
\widetilde{Q}(|k_\perp|) = \frac{i}{k_\perp^2}\int d^2r_\perp e^{ik_\perp \cdot r_\perp} k_\perp \cdot r_\perp Q(|r_\perp|) = -\frac{2\pi}{|k_\perp|}\int_0^\infty  d|r_\perp| r_\perp^2 J_1(|k_\perp| |r_\perp|) Q(|r_\perp|)\,,
\eeq
 is now a real function. Substituting this into (\ref{s}), one gets \cite{Zhou:2013gsa,Boer:2015pni}
\beq
\frac{N_ck_\perp^2M}{2g^2\pi^3}{\cal A} \widetilde{Q}(|k_\perp|) = xf_{1T}^{\perp}(x,k_\perp^2)\,, \label{def}
\eeq
where ${\cal A}$ is the trasnverse area of the nucleon and the superscript $[+-]$ has been omitted.   
In the literature, often the notation $\Delta^N f$  is used instead of $f_{1T}^\perp$. Their relation is 
 (see (2) of  \cite{DAlesio:2015fwo})
 \beq
 \Delta^N f(|k_\perp|) = -2\frac{|k_\perp|}{M} f_{1T}^\perp(|k_\perp|)= - \frac{|k_\perp|^3 N_c}{4\pi^4 \alpha_s x} {\cal A} \widetilde{Q}(|k_\perp|)\,.
 \eeq


\section{Numerical solution of the Odderon equation}
\label{sec:solution}

\subsection{Evolution equation}
\label{sec:evolution}
So far we have kept the $x$-dependence of the Pomeron and Odderon amplitudes implicit. The dependence comes from the small-$x$ evolution which we now turn to. 
The dipole S-matrix satisfies the Balitsky-Kovchegov equation
\beq
\frac{\partial}{\partial Y} S(x_\perp,y_\perp) =\frac{\bar{\alpha}_s}{2\pi} \int \frac{d^2z_\perp}{(2\pi)^2} \frac{(x_\perp-y_\perp)^2}{(x_\perp-z_\perp)^2(z_\perp-y_\perp)^2} (S(x_\perp,z_\perp)S(z_\perp,y_\perp)-S(x_\perp,y_\perp))\,, \label{bk}
\eeq
where $\bar{\alpha}_s\equiv \frac{\alpha_s N_c}{\pi}$ and $Y=\ln \frac{x_0}{x}$ is the rapidity. ($x_0$ is some starting value.)
By decomposing (\ref{bk}) into the real and imaginary parts using (\ref{ji}), one obtains the small-$x$ evolution equations for the Pomeron and Odderon. Let us denote 
\beq
P(x_\perp,y_\perp) = 1-N(x_\perp,y_\perp), \qquad S_\perp \times r_\perp Q(x_\perp,y_\perp) = O(x_\perp,y_\perp).
\eeq
Then the equations read \cite{Kovchegov:2003dm,Hatta:2005as}
\beq
\frac{\partial}{\partial Y} O(x_\perp,y_\perp) &=& \frac{\bar{\alpha}_s}{2\pi} \int d^2z_\perp 
\frac{(x_\perp-y_\perp)^2}{(x_\perp-z_\perp)^2(z_\perp-y_\perp)^2} \Biggl[ O(x_\perp,z_\perp)+O(z_\perp,y_\perp)-O(x_\perp,y_\perp) \nn && \qquad -O(x_\perp,z_\perp)N(z_\perp,y_\perp) -N(x_\perp,z_\perp)O(z_\perp,y_\perp) \Biggr]\,, \label{odd}
\eeq
\beq
\frac{\partial}{\partial Y} N(x_\perp,y_\perp) &=& \frac{\bar{\alpha}_s}{2\pi} \int d^2z_\perp 
\frac{(x_\perp-y_\perp)^2}{(x_\perp-z_\perp)^2(z_\perp-y_\perp)^2} \Biggl[ N(x_\perp,z_\perp)+N(z_\perp,y_\perp)-N(x_\perp,y_\perp) \nn && \qquad -N(x_\perp,z_\perp)N(z_\perp,y_\perp) +O(x_\perp,z_\perp)O(z_\perp,y_\perp) \Biggr]\,. \label{pom}
\eeq
Let $\theta$ be the azimuthal angle in the transverse plane measured with respect to $S_\perp$. Then $O \propto \sin \theta$ and 
the last term $\sim OO$ in (\ref{pom}) induces an angular dependence $\sim \cos 2\theta$ in $N$ even if it is initially independent of $\theta$.\footnote{It should be noted that Ref.~\cite{Kovchegov:2003dm} dismissed this term as a higher order effect which should not be kept consistently to the order of accuracy (i.e., leading logarithmic approximation). However, we keep this term as it brings in qualitatively new interesting features. In any case, the effect of this term is numerically very small.}   This in turn induces a $\sin \theta \cos 2\theta$ dependence in $O$ via (\ref{odd}) (or $\cos 2\theta$ dependence in $Q$). By repeating this argument, $N$ and $Q$ eventually depend on arbitrary even harmonics $\cos 2n \theta$.

The equation for the Odderon has been previously solved numerically in \cite{Lappi:2016gqe}, but the connection to spin physics was not explored.  We shall present our numerical solution of the coupled equation (\ref{odd}) and (\ref{pom}), and study the dependence of the gluon Sivers function on $x$ and $k_\perp$. 

\subsection{Setup}
\label{sec:setup}
Since the Sivers function is a forward matrix element we can assume translational invariance and neglect the dependence of $N$ and $O$ on the impact parameter $b_\perp=\frac{x_\perp+y_\perp}{2}$.  We can then set $y_\perp$ as the origin and switch to the polar coordinates so that $S(x_\perp,y_\perp)=S(x_\perp-y_\perp) = S(r_\perp) = S(|r_\perp|,\theta)$. 
Without loss of generality, we choose the spin vector $S_\perp$ to be along the positive $x$-axis, so $\theta$ is measured with respect to this axis. 

The initial conditions are \cite{Zhou:2013gsa}
\beq
N(|r_\perp|, \theta)=1-e^{-r_\perp^2 Q_{s0}^2}, \qquad O(|r_\perp|,\theta) =  \kappa Q_{s0}^3 |r_\perp|^3\sin\theta e^{-r_\perp^2Q_{s0}^2},
\eeq
at $Y=0$, where $Q_{s0}$ is the initial saturation scale. In principle, the constant $\kappa$ can be determined along the line of \cite{Zhou:2013gsa,Jeon:2005cf},\footnote{There is also a constraint on the value of $\kappa$ from group theory   \cite{Lappi:2016gqe}.}  but here we consider it as an unknown, including its sign. The following numerical results are obtained for  $\kappa = \frac{1}{3}$. Since the evolution equation is essentially linear in $O$ (the nonlinear term $OO$ in (\ref{pom}) is numerically tiny), the normalization and the sign can be adjusted later. It can be also fitted by comparing with the future experimental data at the EIC. 

To calculate the integral over $z_\perp$, we discretize it by setting up a lattice in $|r_\perp|$ and $\theta$. We rescale $|r_\perp|$ by $Q_{s0}$ so the calculation is carried out in a unitless way with $|\tilde{r}_\perp|\equiv |r_\perp| Q_{s0}$. The lattice on $|\tilde{r}_\perp|$ is uniform in $\ln{|\tilde{r}_\perp|}$ from $\ln{|\tilde{r}_\perp}| = -12$ to $4$ with the number of lattice sites $n_{|r_\perp|}=400$. The lattice on $\theta$ is uniform from $0$ to $2\pi$ with the number of sites $n_\theta=100$. We set $\bar{\alpha}_s = \frac{\alpha_s N_c}{\pi} = 0.2$ throughout the paper. We solve the evolution in $Y$ by discretizing $\Delta Y = 0.4$ and use the Runge-Kutta 4th order method. At each evolution step $\Delta Y$, we update the value of $N$ and $O$ at each lattice site. The points of $x_\perp-y_\perp$ and $z_\perp-y_\perp$ are on the lattice, but the points of $x_\perp-z_\perp$ are not in general. So to obtain the values of $N$ and $O$ at the points $x_\perp-z_\perp$, a linear interpretation among the lattice sites is applied.

\subsection{Results}
\label{sec:result}
\begin{figure}
\centering
\includegraphics[height=2.2in]{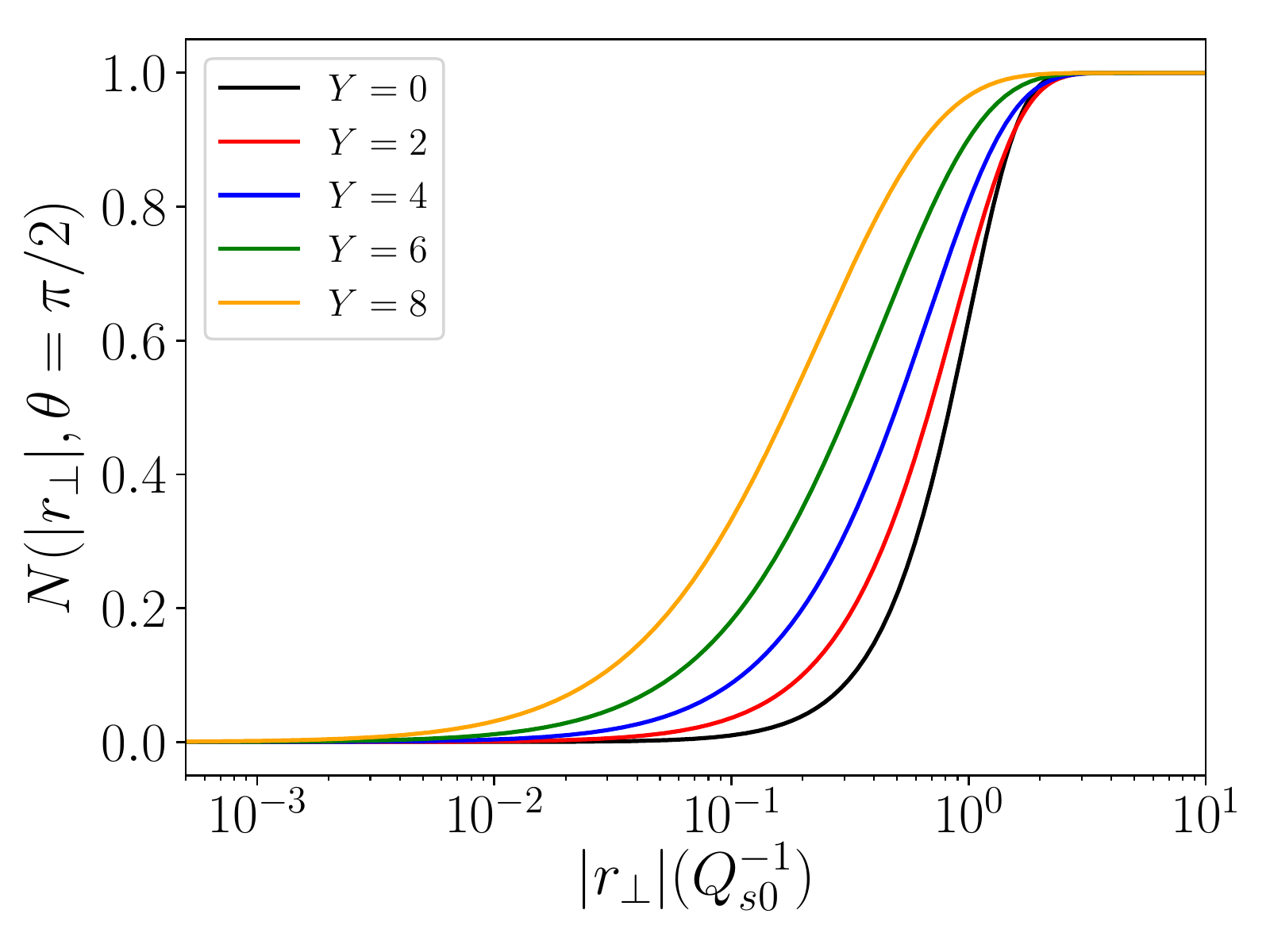}
\includegraphics[height=2.2in]{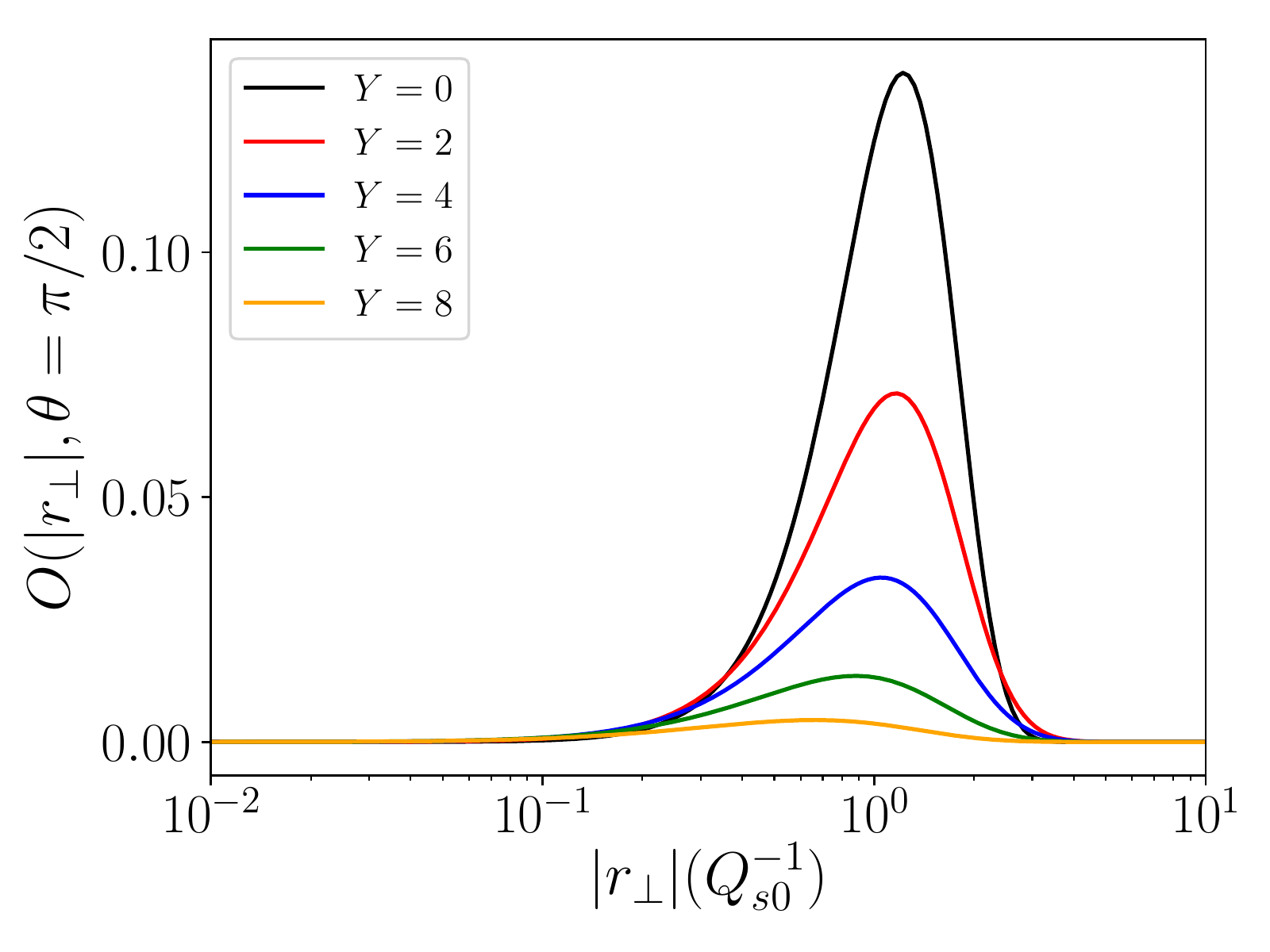}
\caption{$N$ and $O$ as functions of $|r_\perp|$ along the axis of $\theta=\pi/2$.}
\label{fig:NO_vs_r}
\end{figure}

\begin{figure}
\centering
\includegraphics[height=2.2in]{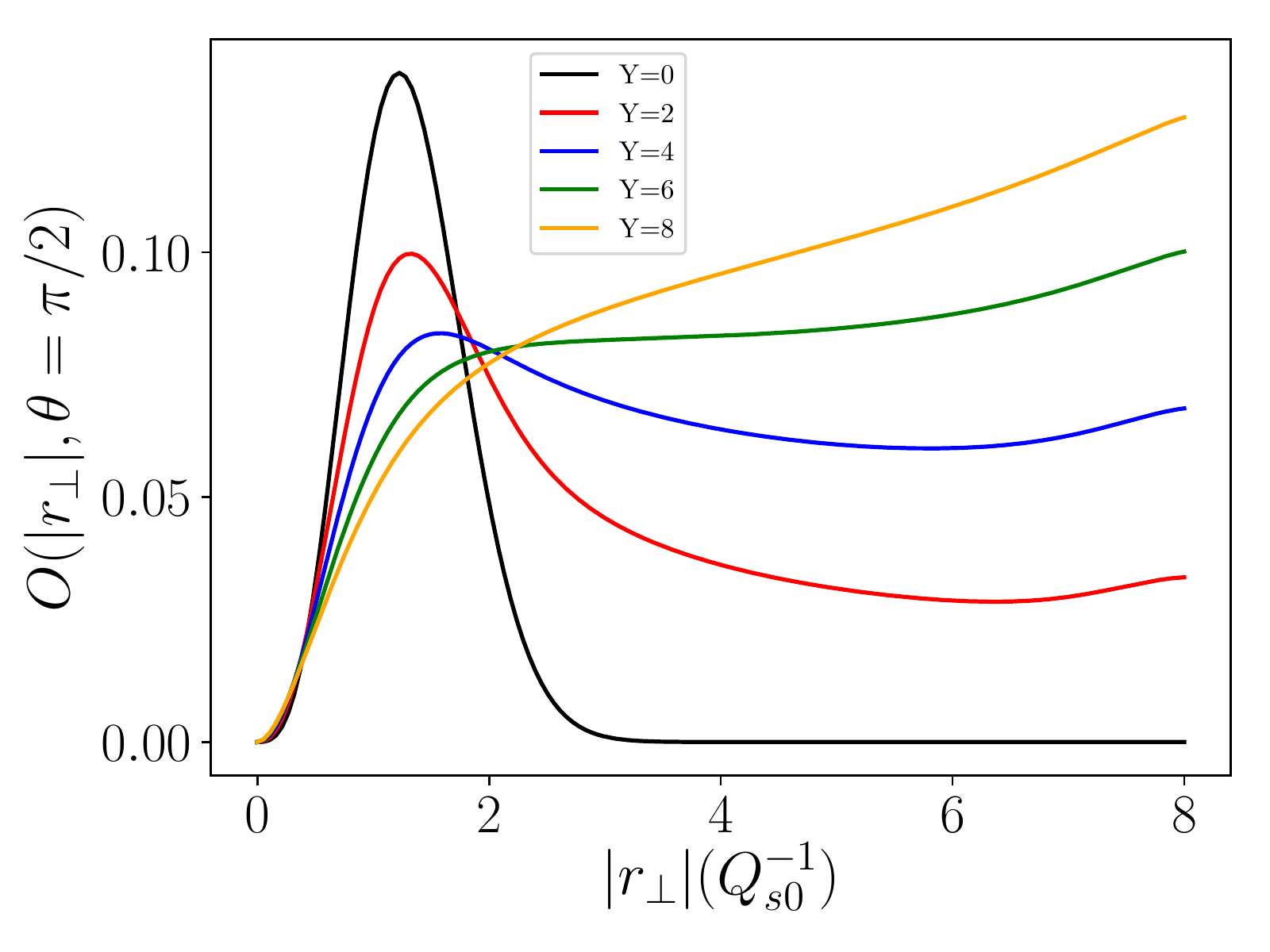}
\caption{ Odderon solution in the linear evolution equation along the axis of $\theta=\pi/2$. 
}
\label{fig:odd_linear}
\end{figure}

We plot the values of $N$ and $O$ at different $Y$'s along the axis of $\theta=\pi/2$ in Fig.~\ref{fig:NO_vs_r}.  The Pomeron amplitude $N$ shows the familiar `geometric scaling' with the inflection point inversely related to the saturation scale $Q_s(Y)\propto e^{\#Y}$. The peak location of the Odderon also moves with $Y$, but its dependence is significantly weaker. This is presumably related to the observation in \cite{Hagiwara:2016kam} that any angular dependence of the dipole S-matrix comes with extra powers of $r_\perp$, and this kills the geometric scaling structure. On the other hand, the magnitude of the Odderon amplitude rapidly decreases with Y. This is due to the nonlinear terms in (\ref{odd}). Indeed, if $N\approx 1$ in the saturation regime, only the virtual term $-O(x_\perp,y_\perp)$ survives on the right hand side of (\ref{odd}), and this leads to a strong Sudakov suppression of $O$. To elaborate on this point, in Fig.~\ref{fig:odd_linear} we show our numerical solution without the nonlinear terms $\sim NO+ON$ (starting from the same initial condition).\footnote{Since we noticed that in this case the large-$r_\perp$ region is not suppressed, we solved the equation on a square lattice $(x,y)$ (not in $\ln \tilde{r}_{\perp}$) with each side from $-8$ to $8$ (scaled by $Q^{-1}_{s0}$) and having $300$ lattice sites.
 }  From the `BLV' Odderon solution 
\cite{Bartels:1999yt}, one expects that the amplitude  is constant in $Y$. However, this is so only in a very limited sense.  We see indeed that the magnitude of $O$ around $|r_\perp|\sim 2Q_{s0}^{-1}$ is almost independent of $Y$, but the smaller- or larger-$r_\perp$ regions strongly depend on $Y$. In the small-$r_\perp$ region the first two terms in (\ref{odd}) nearly cancel   $O(r_\perp-z_\perp)+O(z_\perp) \approx O(-z_\perp) +O(z_\perp)=0$, and therefore the amplitude  receives an exponential Sudakov suppression in $Y$. In the large-$r_\perp$ region, the first and third terms cancel and we are left with the second term $O(z_\perp)$ which is positive. Therefore, the amplitude in the large-$r_\perp$ region tends to increase and becomes sensitive to the finite volume effects as well as the nonperturbative effects.  In this regard it is interesting to see that the nonlinear terms cut off this sensitivity and make the large-$r_\perp$ behavior drastically different.

\begin{figure}
\centering
\includegraphics[height=2.2in]{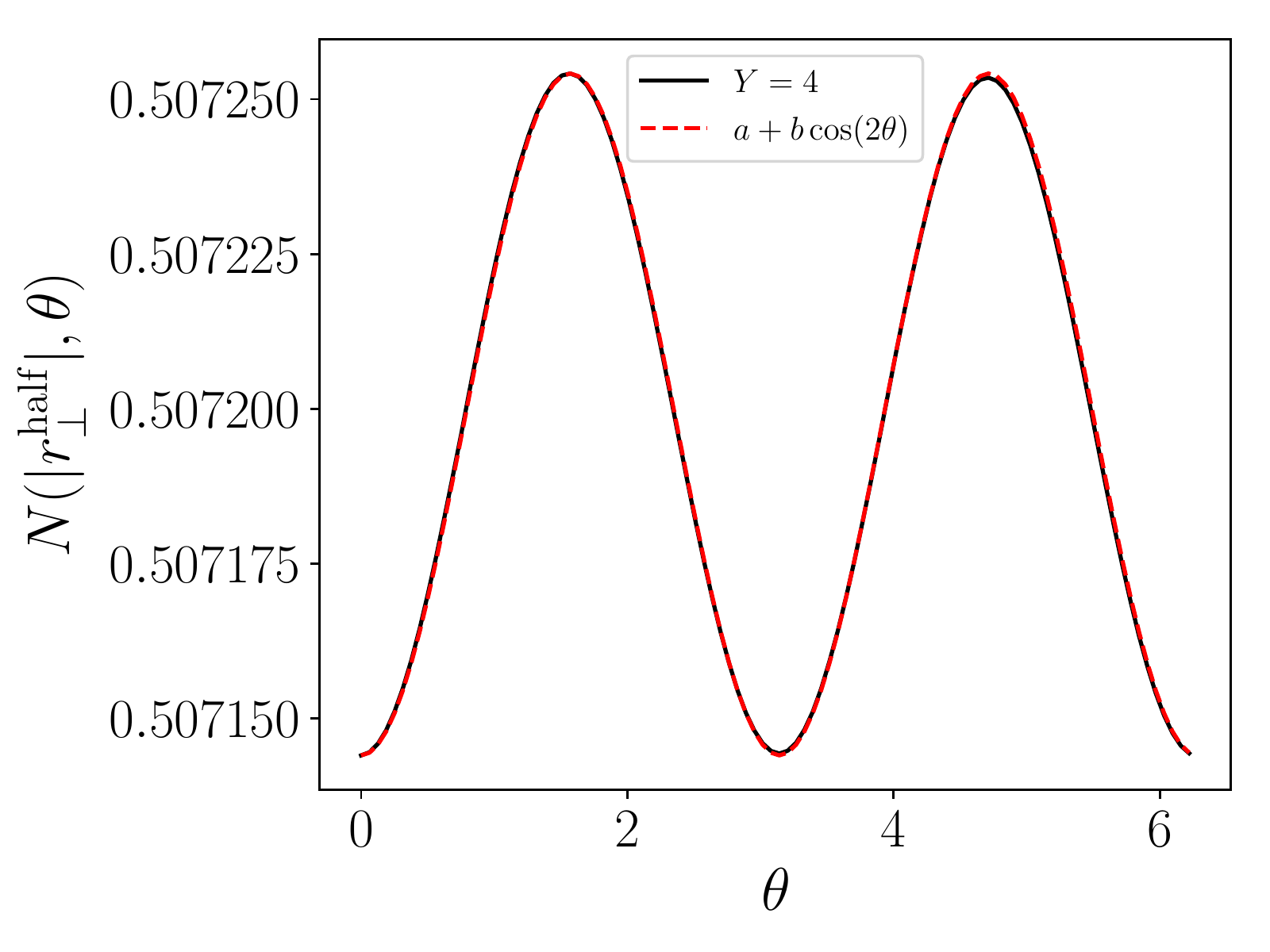}
\caption{$N$ 
as functions of $\theta$ at fixed $|r_\perp|$. $N$ is evaluated at the point $N(|r_\perp^{{\rm half}}|)=0.5$.
The black solid curves represent the numerical solution of the evolution equation while the red dashed curves are fitted results. 
}
\label{fig:NO_vs_theta}
\end{figure}

\begin{figure}
\centering
\includegraphics[height=2.2in]{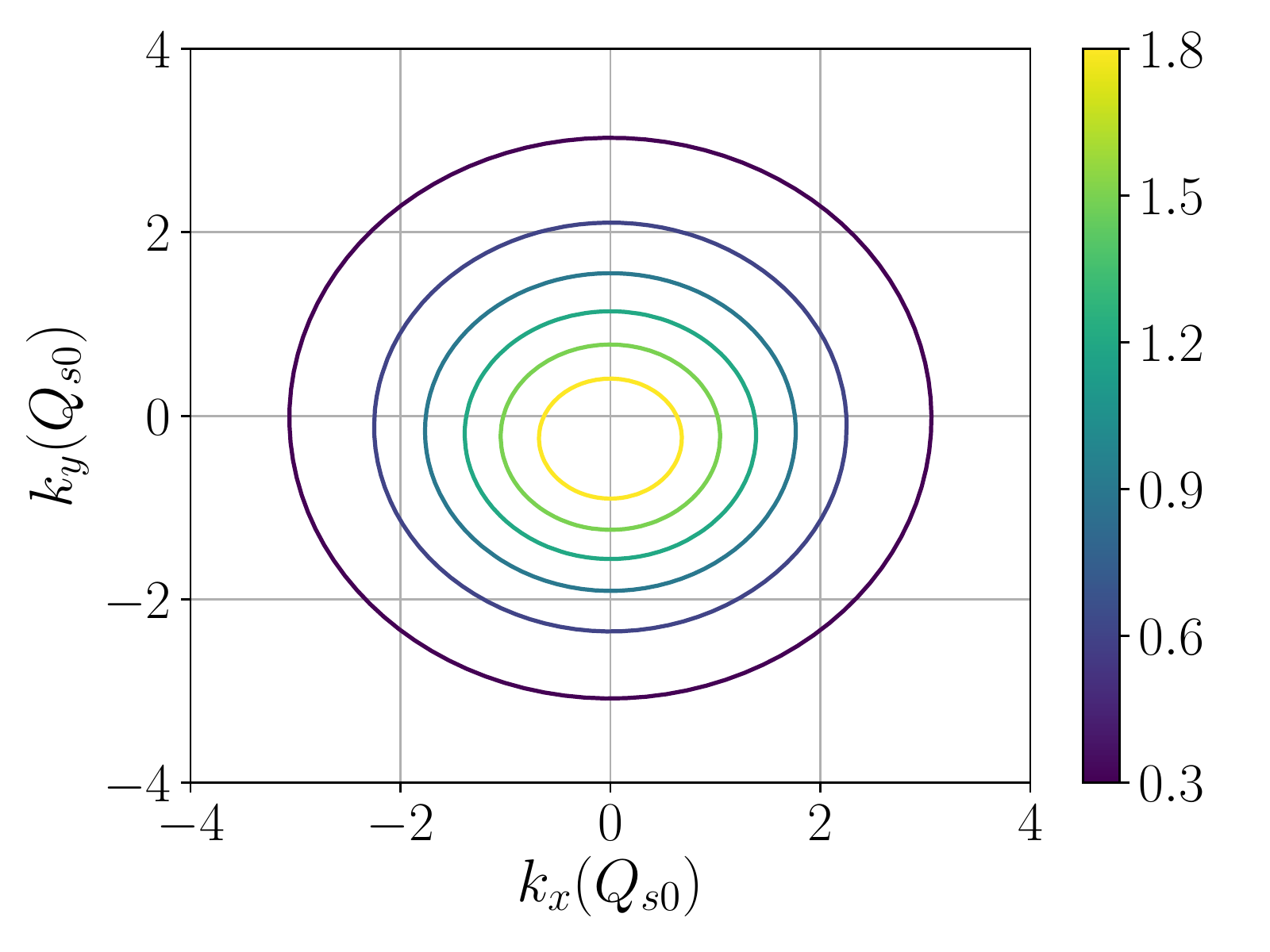}
\caption{Contour plot of (\ref{we}) in the transverse momentum plane at $Y=4$. The spin is pointing in the $+x$ direction.
}
\label{fig:NO_contour}
\end{figure}

Turning to the $\theta$-dependence, in Fig.~\ref{fig:NO_vs_theta} we fix $Y=4$ and $|r_\perp|$ and plot $N$ as a function of $\theta$. We notice that during the evolution in $Y$, the Pomeron $N$ gains a $\cos2\theta$ dependence. This  comes from the $OO$ term in the evolution equation as discussed in Section.~\ref{sec:evolution}.  However, the dependence is extremely weak--the oscillation magnitude in $\theta$ is of order $5\times10^{-5}$. This tiny number is approximately given by $\frac{\bar{\alpha}_s}{2\pi}YO^2$ if we take $Y=4$ and $O\approx0.02$. In principle, $N$ and $O$ will continue influencing each other and develop higher order even harmonics. However, higher order cosine terms are numerically negligible, as can be seen in Fig.~\ref{fig:NO_vs_theta}. 
The contour plot of (\ref{we}) in the $k_\perp$ plane 
at $Y=4$ is shown in Fig.~\ref{fig:NO_contour}. It shows the deformation characteristic to the Sivers function. 

\begin{figure}
\centering
\includegraphics[height=2.2in]{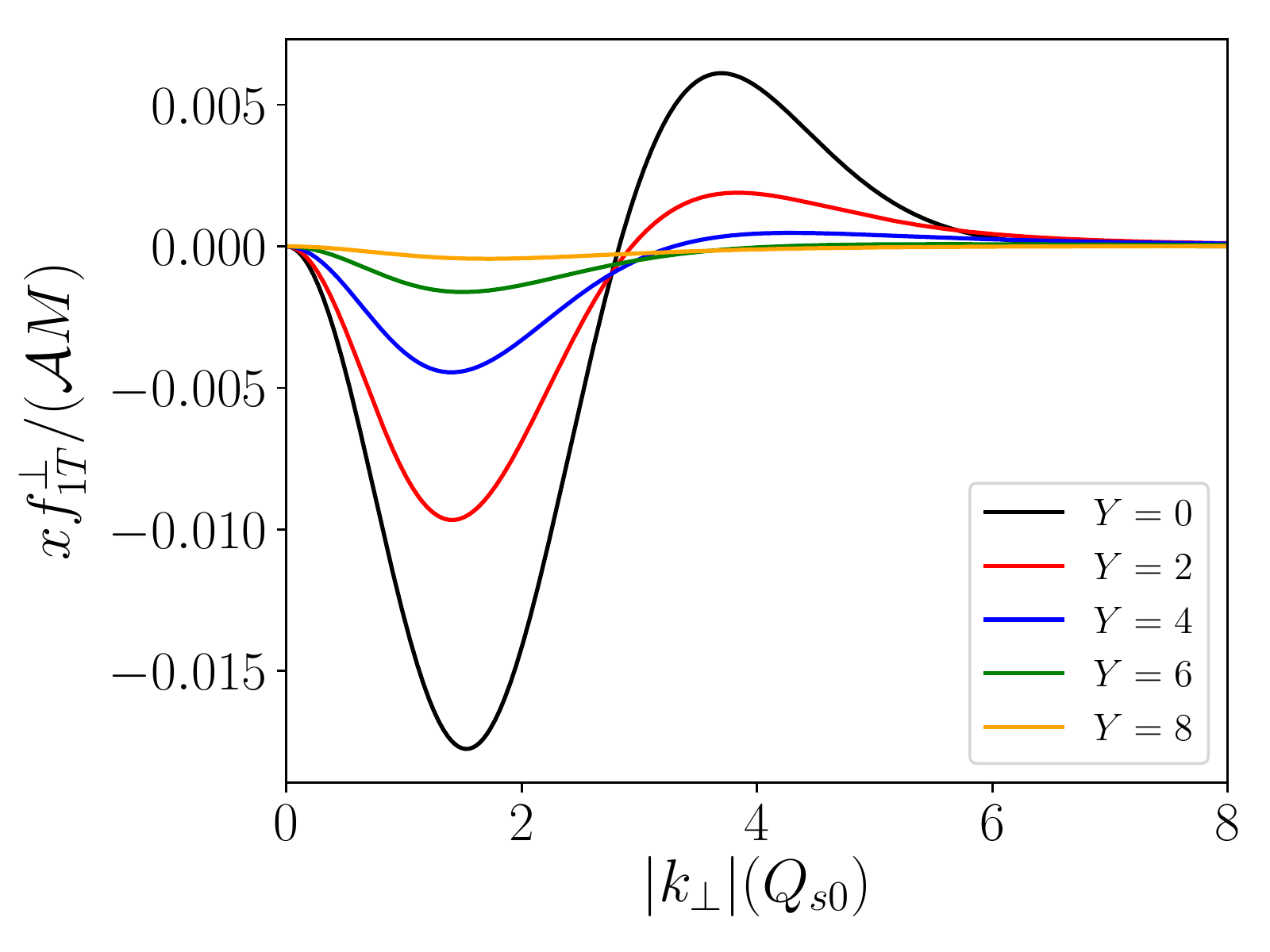}
\caption{ $xf_{1T}^\perp(|k_\perp|)$ at different $Y$'s. The normalization, including the overall sign, is arbitrary  as it depends on the unknown constant $\kappa$. 
}
\label{fig:sivers}
\end{figure}

\begin{figure}
\centering
\includegraphics[height=1.5in]{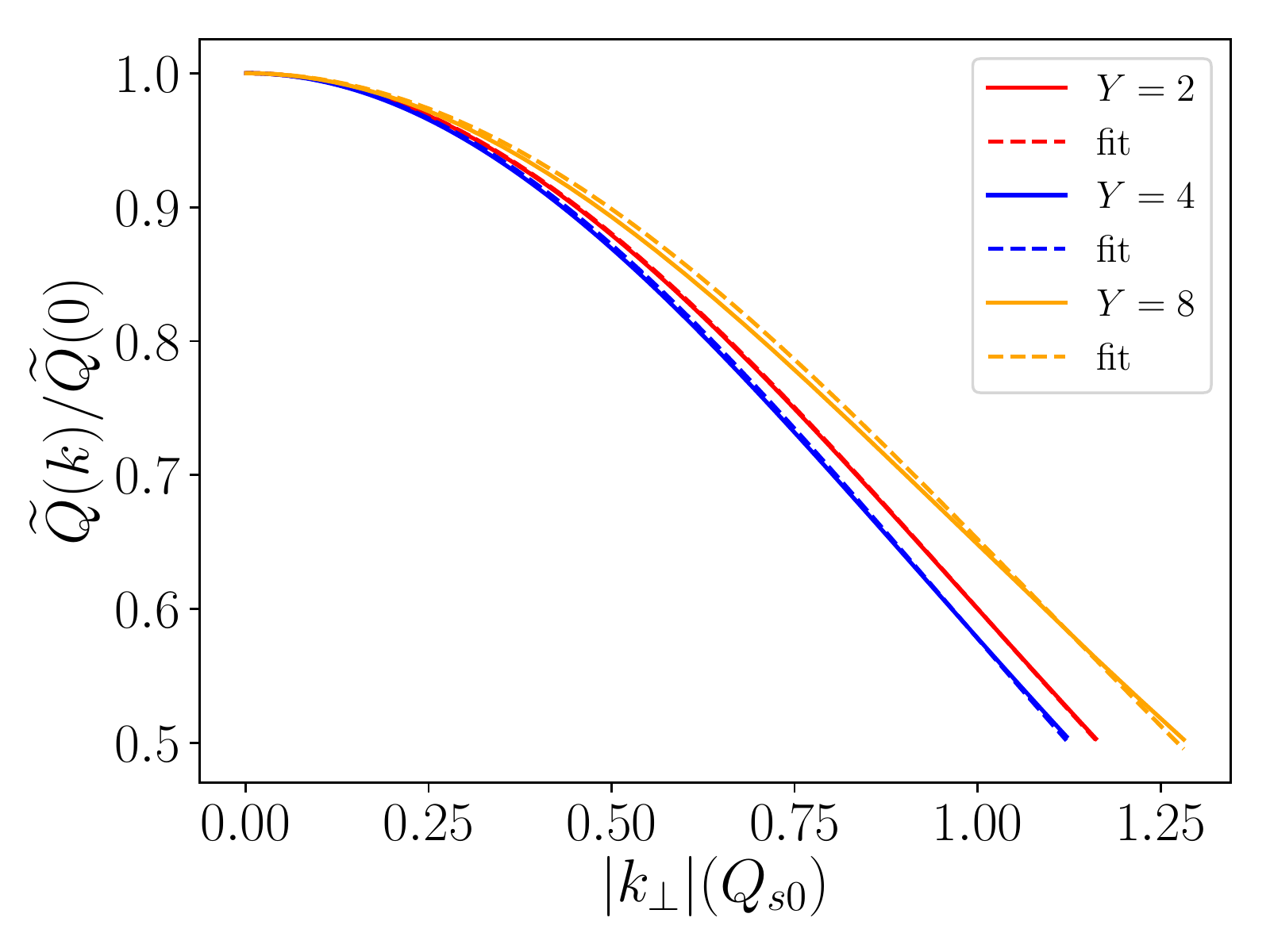}
\includegraphics[height=1.5in]{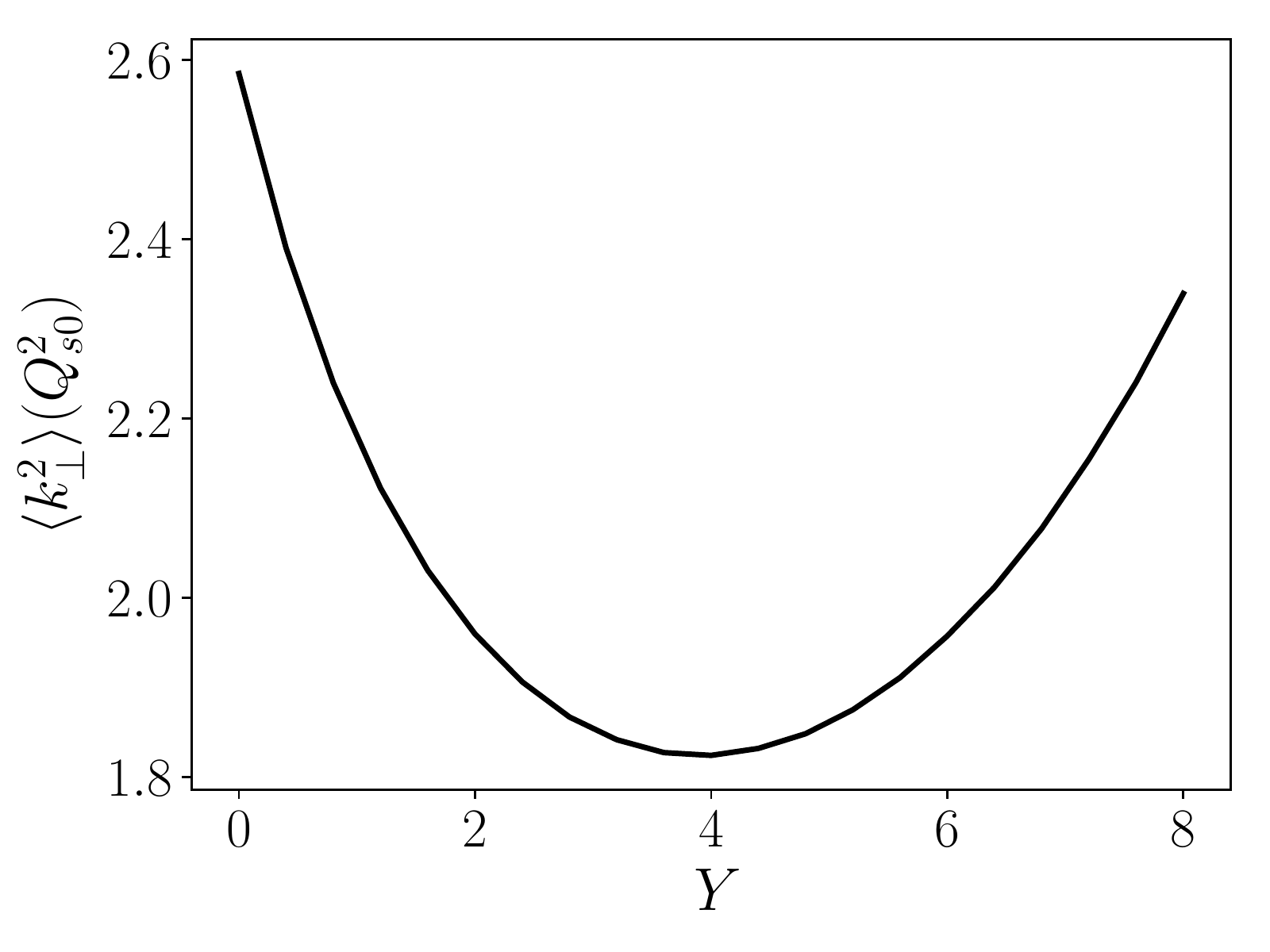}
\caption{Left:  Gaussian fitting of $\widetilde{Q}(k_\perp)/\widetilde{Q}(0)$ (note that $f_{1T}^\perp(k_\perp)\sim k_\perp^2\widetilde{Q}(k_\perp)$, see (\ref{def}). Solid lines are the numerical solutions while dashed lines are the fitting results. Right: the fitted $\langle k_\perp^2 \rangle$ as a function of $Y$.
}
\label{fig:k2}
\end{figure}

The result of the gluon Sivers function is plotted in Fig.~\ref{fig:sivers}. There is a characteristic node for all values of $Y$ which is inherited from the node in $\widetilde{Q}(k_\perp)$ as pointed out in \cite{Zhou:2013gsa}.   We also find that the peak region at small-$k_\perp$ remains approximately Gaussian $\sim |k_\perp|^ne^{-k_\perp^2/\langle k_\perp^2\rangle}$ up to large-$Y$, with some $Y$-dependence in the width $\langle k_\perp\rangle$ as shown in Fig.~\ref{fig:k2}. This means that the $Y=\ln 1/x$ and $k_\perp$ dependences of the Sivers function do not factorize. The fact that   the zero-crossing point in $k_\perp$ depends on $Y$ is also a signal of factorization breaking.
 However,  these effects are rather mild. In the case of unpolarized TMDs, one expects a strong factorization breaking in the small-$x$ regime since both the BFKL and saturation effects induce strong correlations between $x$ and $k_\perp$. In particular, in the latter regime distributions typically depend on the ratio $k_\perp/Q_s(Y)$. However, $\langle k_\perp\rangle$ has a much weaker $Y$-dependence than $Q_s(Y)\sim e^{\# Y}$. As we pointed out already, this is related to the fact that the Odderon is an angular dependent part of the dipole amplitude.  It thus seems that factorization is not a bad approximation for the gluon  Sivers function (and therefore also the sea-quark Sivers function \cite{Dong:2018wsp}).

On the other hand, simple Gaussian parametrizations completely miss the node, and hence the sign change of the Sivers function at large-$k_\perp$. As we shall demonstrate shortly, this sign change has an interesting observable consequence. 
 

\section{SSA in open charm production}

Finally in this section we make a connection to observables.  An interesting process in which the dipole gluon Sivers function plays a role is single spin asymmetry in open charm production $ep^\uparrow \to D^0X,\bar{D}^0X$ in semi-inclusive DIS \cite{Zhou:2013gsa,Godbole:2017fab,Zheng:2018awe}.   
The spin-dependent part of the differential cross section of charm quark production reads
\beq
\frac{d \sigma}{d x_B d z dQ^2 dy d^2\ell_\perp} &=& \frac{\alpha_\ma{em}^2 e_c^2}{2\pi^4x_BQ^2} {\cal A} \int \frac{d^2k_\perp}{(2\pi)^2} H(k_\perp, \ell_\perp, Q^2)   \epsilon^{ij} S_{\perp i} k_{\perp j} \widetilde{Q}(|k_\perp|) ,
\eeq
where 
\beq
H(k_\perp, \ell_\perp, Q^2) &=& \Big[1-y+\frac{y^2}{2}  \Big] \Biggl\{ \big(  z^2+(1-z)^2 \big) \bigg[  \frac{\ell_\perp - k_\perp }{\rho + (\ell_\perp - k_\perp)^2} - \frac{\ell_\perp}{\rho+\ell_\perp^2} \bigg]^2   \nn && \qquad+m_c^2 \left( \frac{1}{\rho+(\ell_\perp -k_\perp)^2}-\frac{1}{\rho+\ell_\perp^2}\right)^2 \Biggr\} \nn 
&&  +4(1-y)z^2(1-z)^2Q^2  \left( \frac{1}{\rho+(\ell_\perp -k_\perp)^2}-\frac{1}{\rho+\ell_\perp^2}\right)^2
\,. \label{jia}
\eeq
The first two lines are contributions from the transversely polarized virtual photon while the last line is the contribution from the longitudinally polarized one.
The kinematic variables are defined as follows: $Q^2 = -q^2$, $x_B=Q^2/2P\cdot q$, $y=P\cdot q/P\cdot k$, $z=P\cdot \ell/P\cdot q$ and $\rho = z(1-z)Q^2+m_c^2$ where $m_c$ is the charm quark mass, and $k$, $P$, $q$ and $\ell$ are the momenta of the incoming electron and proton, the transferred photon and the produced charm quark respectively. Compared to the formula in \cite{Zhou:2013gsa}, we have included the charm quark mass and also added the contribution from the longitudinally polarized virtual photon (cf. \cite{Marquet:2009ca}).

Let us consider the following integral
\beq
\int d^2k_\perp H(k_\perp, \ell_\perp, Q^2) k_{\perp j} \widetilde{Q}(|k_\perp|) \equiv \hat{\ell}_{\perp j}I_c (|\ell_\perp|)\,,
\eeq
where we have used the fact that after integrating over $k_\perp$, the integral should be proportional to $\ell_{\perp j}$. The proportionality constant depends on $|\ell_\perp|$ and can be evaluated straightforwardly using our numerical solution for $\widetilde{Q}(k_\perp)$. Since $H(k_\perp)$ is suppressed when $k_\perp$ is small, we expect that the integral is dominated around $k_\perp\sim \ell_\perp$.  Then an interesting possibility arises that the node in $\widetilde{Q}(k_\perp)$ is transferred to a node in the $\ell_\perp$-spectrum.

\begin{figure}
\centering
\includegraphics[height=1.5in]{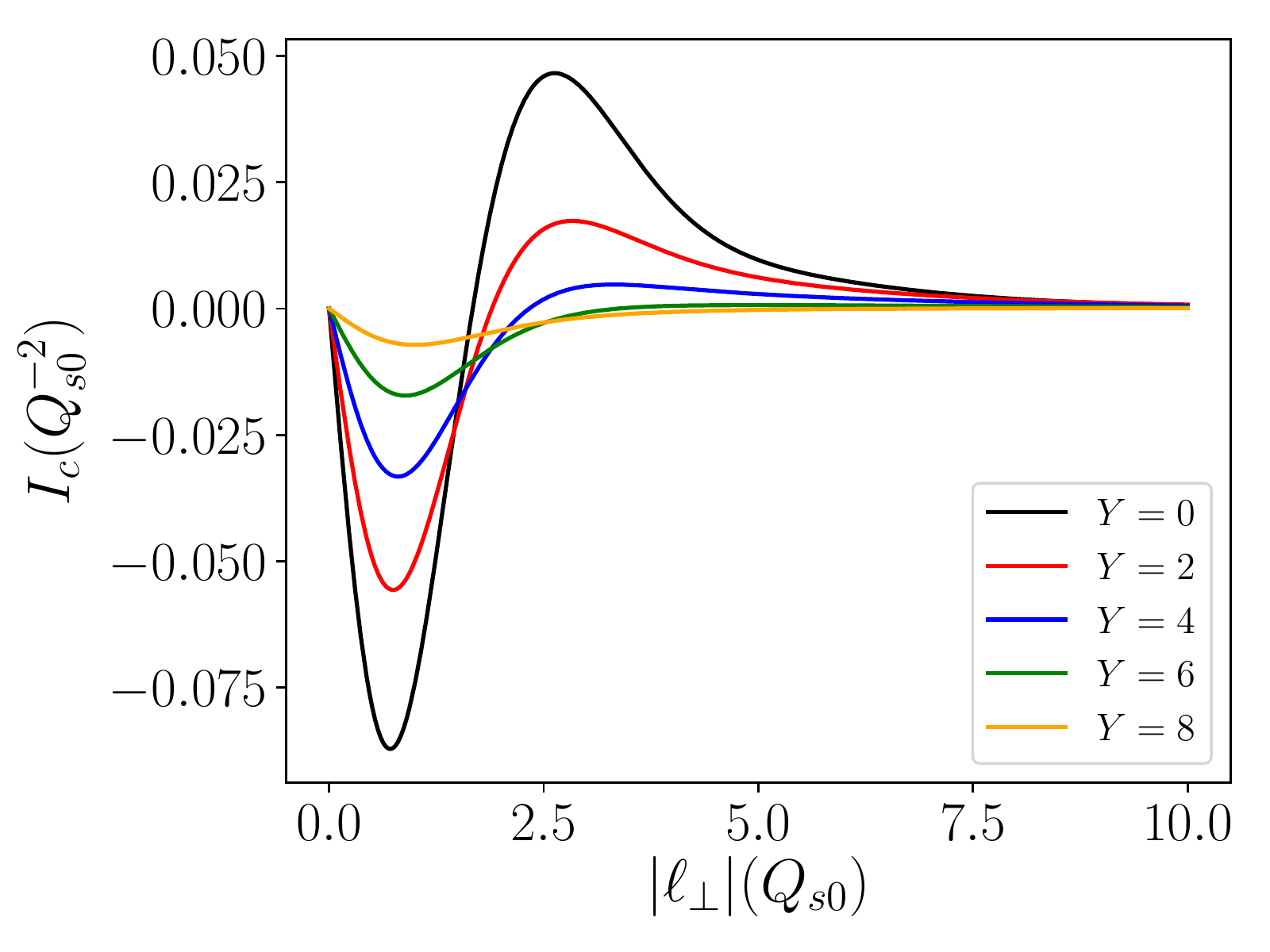}
\includegraphics[height=1.5in]{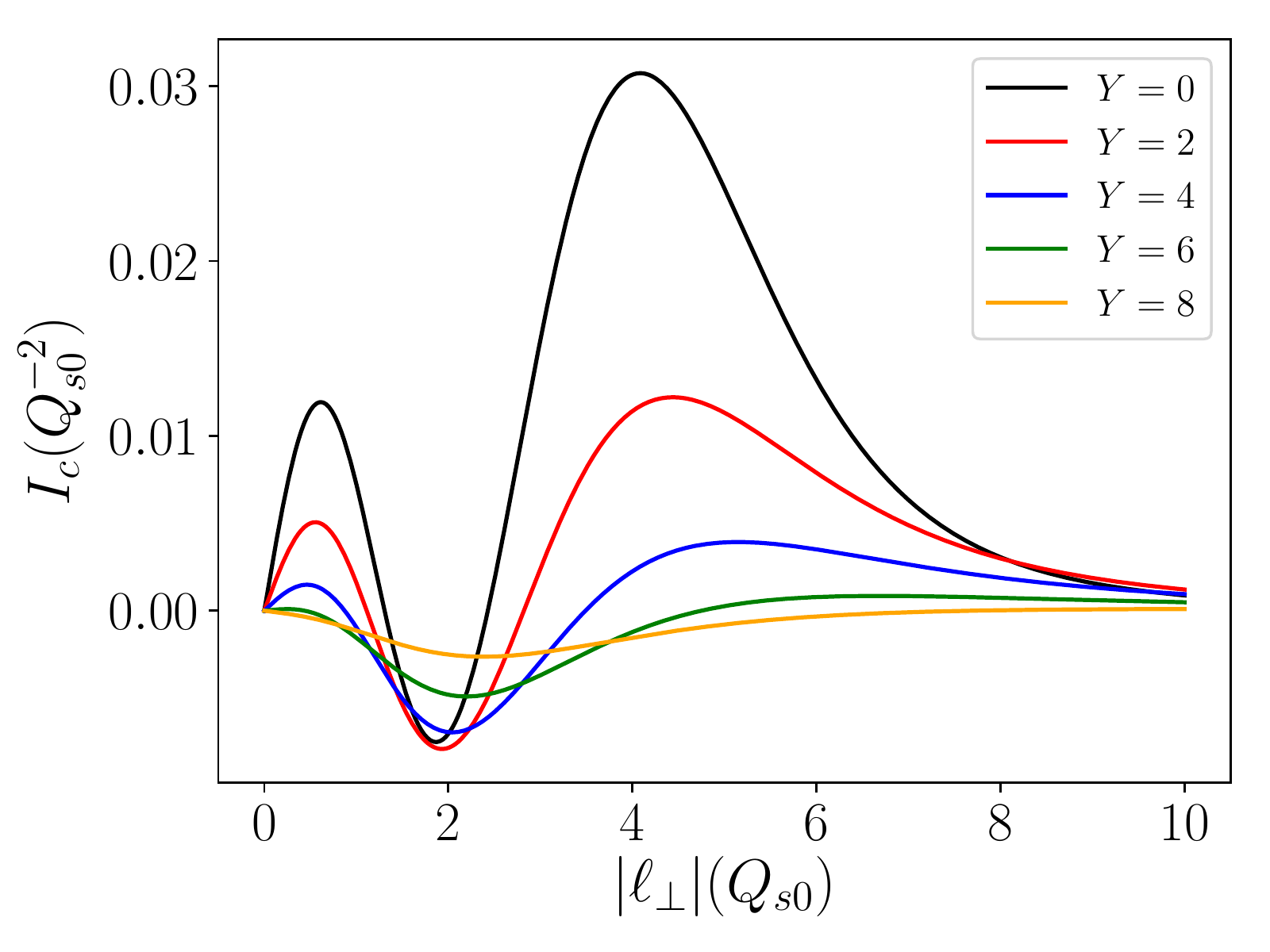}
\includegraphics[height=1.5in]{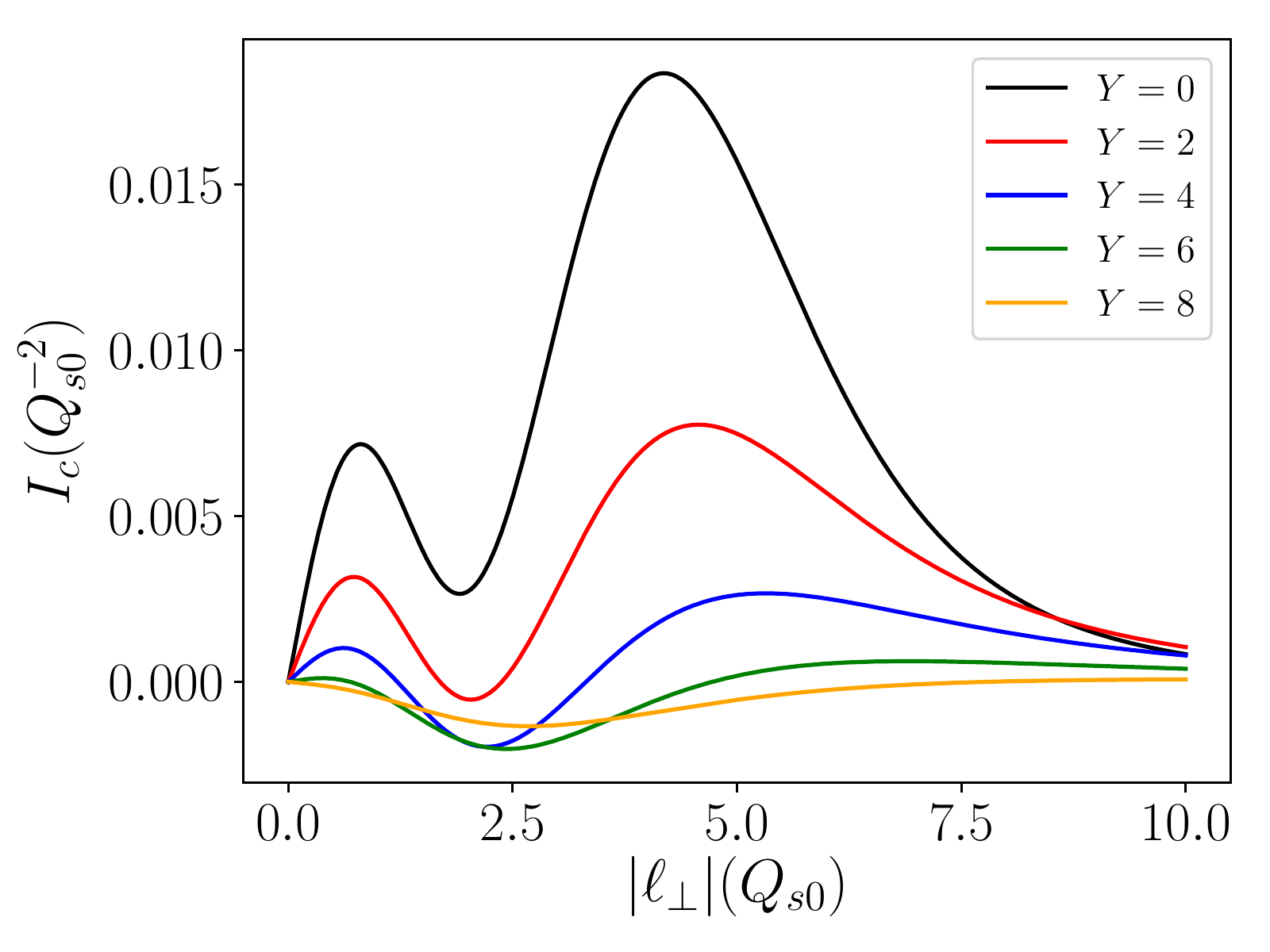}
\caption{ Left: $I_c(|\ell_\perp|)$ from the transversely polarized virtual photon in the massless quark limit at $Q^2 = 5$ GeV. Middle: $I_c(|\ell_\perp|)$ at $Q^2 = 1$ GeV. Right: $I_c(|\ell_\perp|)$ at $Q^2 = 5$ GeV. }
\label{fig:Ic}
\end{figure}

We assume $m_c=1.3$ GeV, $Q_{s0} = 0.7$ GeV and evaluate $I_c$ for $y=z=1/2$ and different values of $Q^2$. We first plot in Fig.~\ref{fig:Ic} (left) the contribution coming only from the first line of (\ref{jia}). As expected, we see very clearly a sign change roughly around the same value of transverse momentum where $\widetilde{Q}$ changes signs.  However, the situation becomes more complicated once the other terms are included. Actually, unless $Q^2$ is very large, the middle term of (\ref{jia}) is always dominant  and this flips the sign of $I_c$ in the low-$\ell_\perp$ region. As a result, the SSA flips signs twice as the transverse momentum of the produced charm quark is varied. This is shown in the middle plot of Fig.~\ref{fig:Ic}. The signal gradually decreases and the sign flipping almost disappears with increasing $Q^2$.\footnote{This is consistent with the TMD evolution effects which tend to diminish the node structure at high-$Q^2$ \cite{Dong:2018wsp}.} But the valley structure in the SSA as a function of transverse momentum still exists. It is thus very interesting to vary $Q^2$ in experiments and explore these different possibilities. 

\section{Conclusions}
\label{sec:conclusion}

In this paper we have computed the (dipole) gluon Sivers function at small-$x$ by numerically solving the evolution equation for the Odderon \cite{Kovchegov:2003dm,Hatta:2005as}. 
Our results convey several important messages which should be considered when modeling the Sivers function. Firstly, the factorization breaking of the $x$ and $k_\perp$-dependences of the Sivers function is rather mild, in contrast to the unpolarized TMDs which is related to the Pomeron part of the dipole amplitude. This seems to be a generic feature of the angular dependent part of the Wilson line correlators (cf. \cite{Hagiwara:2016kam}), and thus we expect that the situation is similar for the Weisz$\ddot{{\rm a}}$cker-Williams gluon Sivers function. In other words, assuming factorization may not be a bad approximation, although systematic deviations do exist.  Secondly, the amplitude of the Sivers function decreases monotonically in $Y$ in the large-$k_\perp$ (small-$r_\perp$) region in the linear evolution regime, and everywhere in the nonlinear regime. This is due to the dominance of the virtual term $-O(x_\perp,y_\perp)$ in the evolution equations which leads to an exponential suppression.  Finally, the dipole gluon Sivers function has a node \cite{Zhou:2013gsa} for all values of $x$. This leads to interesting observable consequences: the SSA in open charm production $ep^\uparrow\to DX$ in SIDIS changes signs, possibly twice, as a function of the D-meson transverse momentum at low-$Q^2$. Note also that SSA for $D^0$ mesons has an opposite sign from that for $\bar{D}^0$ mesons due to the C-odd nature of the Odderon \cite{Beppu:2010qn,Zhou:2013gsa,Kang:2008qh}. On top of this, there is the `usual' sign change in the Weisz$\ddot{{\rm a}}$cker-Williams gluon Sivers function that can be tested in $ep^\uparrow$ and $pp^\uparrow$ collisions  \cite{Boer:2016fqd}. All these make the studies of the gluon Sivers function very fascinating, and we look forward to a rich phenomenology at experimental facilities such as RHIC, AFTER@LHC and EIC. 

\begin{acknowledgments}
We are grateful to Jian Zhou for discussions and for reading the draft of this paper.  We also thank Piet Mulders and Feng Yuan for discussion. This material is based on work supported by the U.~S. Department of Energy, Office of Science, Office of Nuclear Physics under Contract No. DE-SC0012704. It is also supported in part by a LDRD grant from Brookhaven Science Associates. 
X.Y. thanks the nuclear theory group at Brookhaven National Laboratory, where this work was completed, for its hospitality. X.Y. is supported by U.S. Department of Energy under Research Grant No. DE-FG02-05ER41367. X.Y. also acknowledges support from Brookhaven National Laboratory. Y. Hagiwara is supported by the JSPS KAKENHI Grant No. 17J08072.

\end{acknowledgments}


\begin{thebibliography}{99}

\bibitem{Sivers:1989cc} 
  D.~W.~Sivers,
  Phys.\ Rev.\ D {\bf 41}, 83 (1990).

\bibitem{Collins:2002kn} 
  J.~C.~Collins,
  Phys.\ Lett.\ B {\bf 536}, 43 (2002)
  [hep-ph/0204004].

\bibitem{Adamczyk:2015gyk} 
  L.~Adamczyk {\it et al.} [STAR Collaboration],
  Phys.\ Rev.\ Lett.\  {\bf 116}, no. 13, 132301 (2016)
  [arXiv:1511.06003 [nucl-ex]].

\bibitem{Aghasyan:2017jop} 
  M.~Aghasyan {\it et al.} [COMPASS Collaboration],
  Phys.\ Rev.\ Lett.\  {\bf 119}, no. 11, 112002 (2017)
  [arXiv:1704.00488 [hep-ex]].

\bibitem{Anselmino:2004nk} 
  M.~Anselmino, M.~Boglione, U.~D'Alesio, E.~Leader and F.~Murgia,
  Phys.\ Rev.\ D {\bf 70}, 074025 (2004)
  [hep-ph/0407100].

\bibitem{Yuan:2008vn} 
  F.~Yuan,
  Phys.\ Rev.\ D {\bf 78}, 014024 (2008)
  [arXiv:0801.4357 [hep-ph]].


\bibitem{Zhou:2013gsa} 
  J.~Zhou,
  Phys.\ Rev.\ D {\bf 89}, no. 7, 074050 (2014)
  [arXiv:1308.5912 [hep-ph]].

\bibitem{DAlesio:2015fwo} 
  U.~D'Alesio, F.~Murgia and C.~Pisano,
  JHEP {\bf 1509}, 119 (2015)
  [arXiv:1506.03078 [hep-ph]].


\bibitem{Boer:2015pni} 
  D.~Boer, M.~G.~Echevarria, P.~Mulders and J.~Zhou,
  Phys.\ Rev.\ Lett.\  {\bf 116}, no. 12, 122001 (2016)
  [arXiv:1511.03485 [hep-ph]].


\bibitem{Boer:2016fqd} 
  D.~Boer, P.~J.~Mulders, C.~Pisano and J.~Zhou,
  JHEP {\bf 1608}, 001 (2016)
  [arXiv:1605.07934 [hep-ph]].



\bibitem{Godbole:2017fab} 
  R.~M.~Godbole, A.~Kaushik and A.~Misra,
  Phys.\ Rev.\ D {\bf 97}, no. 7, 076001 (2018)
  [arXiv:1709.03074 [hep-ph]].


\bibitem{Goncalves:2017fkt} 
  V.~P.~Goncalves,
  Phys.\ Rev.\ D {\bf 97}, no. 1, 014001 (2018)
  [arXiv:1710.01674 [hep-ph]].

\bibitem{Zheng:2018awe} 
  L.~Zheng, E.~C.~Aschenauer, J.~H.~Lee, B.~W.~Xiao and Z.~B.~Yin,
  Phys.\ Rev.\ D {\bf 98}, no. 3, 034011 (2018)
  [arXiv:1805.05290 [hep-ph]].

\bibitem{Hadjidakis:2018ifr} 
  C.~Hadjidakis {\it et al.},
  arXiv:1807.00603 [hep-ex].

\bibitem{Rajesh:2018qks} 
  S.~Rajesh, R.~Kishore and A.~Mukherjee,
  Phys.\ Rev.\ D {\bf 98}, no. 1, 014007 (2018)
  [arXiv:1802.10359 [hep-ph]].

\bibitem{Dong:2018wsp} 
  H.~Dong, D.~X.~Zheng and J.~Zhou,
  arXiv:1805.09479 [hep-ph].





\bibitem{Godbole:2018mmh} 
  R.~M.~Godbole, A.~Kaushik, A.~Misra and S.~Padval,
  arXiv:1810.07113 [hep-ph].

\bibitem{DAlesio:2018rnv} 
  U.~D'Alesio, C.~Flore, F.~Murgia, C.~Pisano and P.~Taels,
  arXiv:1811.02970 [hep-ph].


\bibitem{Boer:2015vso} 
  D.~Boer, C.~Lorce, C.~Pisano and J.~Zhou,
  Adv.\ High Energy Phys.\  {\bf 2015}, 371396 (2015)
  [arXiv:1504.04332 [hep-ph]].






\bibitem{Szymanowski:2016mbq} 
  L.~Szymanowski and J.~Zhou,
  Phys.\ Lett.\ B {\bf 760}, 249 (2016)
  [arXiv:1604.03207 [hep-ph]].


\bibitem{Bartels:1999yt} 
  J.~Bartels, L.~N.~Lipatov and G.~P.~Vacca,
  Phys.\ Lett.\ B {\bf 477}, 178 (2000)
  [hep-ph/9912423].


\bibitem{Kovchegov:2003dm} 
  Y.~V.~Kovchegov, L.~Szymanowski and S.~Wallon,
  Phys.\ Lett.\ B {\bf 586}, 267 (2004)
  [hep-ph/0309281].



\bibitem{Hatta:2005as} 
  Y.~Hatta, E.~Iancu, K.~Itakura and L.~McLerran,
  Nucl.\ Phys.\ A {\bf 760}, 172 (2005)
  [hep-ph/0501171].

\bibitem{Lappi:2016gqe} 
  T.~Lappi, A.~Ramnath, K.~Rummukainen and H.~Weigert,
  Phys.\ Rev.\ D {\bf 94}, no. 5, 054014 (2016)
  [arXiv:1606.00551 [hep-ph]].

\bibitem{Jeon:2005cf} 
  S.~Jeon and R.~Venugopalan,
  Phys.\ Rev.\ D {\bf 71}, 125003 (2005)
  [hep-ph/0503219].


\bibitem{Kovchegov:2012ga} 
  Y.~V.~Kovchegov and M.~D.~Sievert,
  Phys.\ Rev.\ D {\bf 86}, 034028 (2012)
  Erratum: [Phys.\ Rev.\ D {\bf 86}, 079906 (2012)]
  [arXiv:1201.5890 [hep-ph]].



\bibitem{Hagiwara:2016kam} 
  Y.~Hagiwara, Y.~Hatta and T.~Ueda,
  Phys.\ Rev.\ D {\bf 94}, no. 9, 094036 (2016)
  [arXiv:1609.05773 [hep-ph]].



\bibitem{Marquet:2009ca} 
  C.~Marquet, B.~W.~Xiao and F.~Yuan,
  Phys.\ Lett.\ B {\bf 682}, 207 (2009)
  [arXiv:0906.1454 [hep-ph]].

\bibitem{Beppu:2010qn} 
  H.~Beppu, Y.~Koike, K.~Tanaka and S.~Yoshida,
  Phys.\ Rev.\ D {\bf 82}, 054005 (2010)
  [arXiv:1007.2034 [hep-ph]].

\bibitem{Kang:2008qh} 
  Z.~B.~Kang and J.~W.~Qiu,
  Phys.\ Rev.\ D {\bf 78}, 034005 (2008)
  [arXiv:0806.1970 [hep-ph]].

\end{thebibliography}
\end{document}